# Towards an augmented reality fourth generation social networks

Andrés Montero

*The Concept Foundation, Madrid, Spain*

Borja Belaza

*Conception Singularity Innovations, Madrid, Spain*

contact@conception.is

**Abstract**

*A concept of fourth generation social network is described as one that, built on the features of augmented reality (AR), is able to implement an enriched layer of digital information that displays in People Augmented Reality (PAR) devices data shared by users in social networks. This PAR layer is accessed by the users in their devices through camera effects when targeting with a mobile phone to a user holding a mobile device with AGPS and with a profile in social media. The social network of fourth generation will be a combination between Facebook and Pokemon Go.*

## I. INTRODUCTION

Currently the concept of social networks is directly associated with social media based on the World Wide Web, such as Twitter, Facebook, Google+, Qzone (China) or VKontakte (Russia), among others.

However, the seed of social networks, which could be considered the *first generation* of social networks, was in the inter-connection of users through terminal emulators (tty) in discussion groups, bulletin board systems (BBS) and electronic mailing lists (eg LISTSERV) of the late 1970s and early 1980s in the 20$^{th}$ century. All of these systems of connection between individuals were computer-mediated communications (CMC) where users interacted with each other via early modems based on acoustic couplers to handle analogical telephone operations [1]. The interaction was mainly developed in text mode and was displayed on monochrome monitors in the pre-internet era; in fact, one of those early first generation social networks was ARPANET, considered a precursor of the Internet due to its implementation of the TCP/IP protocol in 1982.

The *second generation* of social networks could be considered those that, based on the previous ones, move to reside in the Word Wide Web and adopt the protocol TCP/IP for communications and smartmodems to get connected. Tripod, GeoCities or TheGlobe were the clearest representatives of this second wave of social networks that did not have the epicentre in the individual but in the creation of online communities where individuals met to exchange information and discuss various topics and interests, adopting usually the format of a forum or an online chat.

The *third generation* of social networks appeared when the community of shared interests on the web was replaced by the individual profiles that constitute their own communities in the form of groups of friends. This third generation took benefit from technological advances in hardware, software and communications, such as the expansion of broadband (asymmetric digital subscriber line, ADSL) modems or smartphones, thus social networks becoming mobile. The first of these third generation social networks came up in the late nineties of the 20$^{th}$ century with SixDegrees, Makeoutclub or Friendster, followed quickly by the best known MySpace, Linkedin, Orkut or 4Chan, to give way for Facebook in 2004.

Leaving aside technological innovations in terms of telecommunication from which undoubtedly all current social networks have been fed, from a strictly functional perspective the defining characteristics of third generation social networks are in their user-centrality: they connect personal profiles through a website or a mobile app; are designed for the mass dissemination of user-generated content (UGC) in any format; and are moving forward to implement ways of instant communication between users such as instant messaging or video calling.

## II. DIGITAL '2D' REALITY

Since its first generation social networks have been mainly vehicles for the connection between individuals and the transmission of information in a two dimensional virtual space.

The social media contents are generated (or captured in audio, image or video) by the users and shared in a web or in a mobile app. The content resides in a virtual two-dimensional space (web or mobile app) that the users visit telematically to interact with those contents or other users.

The contents shared in social networks are therefore two-dimensional representations displayed in a screen of three-dimensional realities in the physical world: photographs or videos of the user himself, of buildings and places, along with comments in text that represent thoughts, feelings or behaviours of people, along with the transmission of ideas or interests representing collectives. This set creates a digital reality (DR) in the cyber universe which is a two-dimensional (plane) representation of analogical reality in the three dimensional physical universe (volume).

## III. VIRTUAL VS. AUGMENTED '3D' REALITIES

Although cyberspace is currently a concept with many meanings several of them related to the field of military defence [2], it is often used as an Internet metaphor to describe the presence of objects and identities that "exist in cyberspace" through contents located and shared on the Internet [3].

In this last metaphorical sense, social networks in particular would be understood as *territories* within that cyberspace where identities and objects of the analogical world represented by means of two-dimensional digital contents in the form of text, images, audios and videos would have a *residence*. However, even social networks (or websites of various types without the features of a social network) can represent objects or identities without any correspondence with an analogical reality of the physical world: they can be simply imagined or fictitious digital realities, created by users with different intentionality.

Thus, social networks represent the physical world digitally but are not the physical world. In terms of mathematical set theory, the correspondence between physical analogical reality and digital reality of social networks would be an injective function without properties of morphism (f: $R^3 \rightarrow DR^2$). That is, the digital representation ($DR^2$) of the analogical reality of the physical world ($R^3$) provided by third-generation social media does not preserve the three-dimensional structure, albeit digitally, of the represented reality.

In contrast to virtual (VR) and augmented (RA) realities, what could be called *two-dimensional digital reality* ($DR^2$) of third-generation social networks does not have the vocation to embody, nor even from a representational point of view, the three-dimensional reality of the analogical physical space.

This vocation to digitize analogical reality by embedding it in a representational three-dimensional space is being developed in parallel by two differentiated technological lines, which are nevertheless looking for convergence: augmented reality (AR) and virtual reality (VR).

Augmented reality (AR) is an enriched digital version of analogical physical reality overlapping by technological means a layer of digital information. The aim of AR is to inject computer-generated sensory input (geo-location data, text, audio, video, graphics) into a two-dimensional plane of analogical reality displayed through an interface, usually the screen of a smartphone mobile, tablet, computer or other device (glasses, for example). The most popular device of AR in the recent years has been the Google Glass [4].

Within the very same formalism of set theory, augmented reality could be equated with a surjective function of the analogical reality of the physical world ($R^3$) onto an intersection of a layer of two-dimensional digital reality ($DR^2$) and a $R^2$ representation displayed on a screen of the analogical reality of the physical world ($R^3$). In this sense, it would be said that it performs the inverse path to the traditional function of general representativity of the Internet or social networks: instead of digitally representing the physical world, augmented reality creates digital contents that it projects on a two-dimensional plane of the physical world (f: $R^3 \rightarrow DR^2 \cap R^2$).

Regarding virtual reality (VR), it is a computer-generated digital representation of an environment that can be a bijective correspondence of analogical reality, a modified recreation or directly an imaginary setting. The main feature of VR is its three-dimensionality (3D). In order for this 3D to be perceived by an user some kind of VR tools are used as interfaces, being currently a headset connected to a smartphone the most widespread device.

Therefore, the VR takes the user to a three-dimensional representation that may or may not correspond to the analogical physical universe. In the formalization of set theory, the VR could even be a 3D bijective isomorphism between analogical and digital reality (f: $R^3 \Leftrightarrow VR^3$).

The combination of augmented and virtual reality could lead to a homomorphism from analogical reality in $R^3$ to a digital reality in $VR^3$, further enriched with an additional layer of topologies in $AR^2$: the digital content added by augmented reality to the digital representation of the analogical reality (f: $R^3 \rightarrow AR^2 \cap VR^3$).

The main difference between AR and VR as it relates to the user experience is that AR projects digital content over a digital representation of physical reality and the user can access an augmented reality by targeting his/her mobile device over a geo-positioned coordinate of three-dimensional physical reality, displaying the augmented reality on the 2D screen of the device. That is, the user "stay digitally out" of the augmented reality and accedes to it as a "participant viewer". In contrast, in VR the user "enters" into digital reality as a "digital actor" by paying the price of carrying a specific audiovisual device, usually a combination of a headset and a handset connected to a smartphone, sometimes with manual devices for remote control of movement.

Until the arrival of Facebook Spaces in the early 2017 [5], the VR was somehow out to social networks (for instance, 'Second Life' is a virtual world community) and still remains so the AR as a feature in the structural capabilities deployed by the main social media platform for their users. Facebook is introducing AR into its social network through the recently launched 'Camera Effects Platform', whereby developers will be able to use AR Studio to inject 2D digital content on the representation of the analogical physical reality captured by the camera of a mobile device.

In social networks, an AR ersartz was limited to creating digital effects or frames that users could overlap to images or videos, features that have been available for a quite while in Snapchat, Instagram or WhatsApp. With the new AR Studio platform Facebook provide the developers with the capabilities to "code against the real world; to create experiences that are responsive to the environment around you"[6].

Therefore, while the VR connects social networks with a virtualized 3D environment through head-mounted display devices, the AR adds a digital layer to analogical physical reality through the camera of the mobile device, without the need for the user to "mount" other devices on himself, not even some glasses that could operate as interface.

## IV. PEOPLE AUGMENTED REALITY

Although through AR Studio Facebook is proposing to developers to inject a layer of AR in all user experiences with the analogical reality of the physical universe, the real potential revolution of AR in user interaction with a digital representation ($DR^2$) of analogical reality ($R^3$) arrived in July 2016 with an augmented reality game, Pokemon Go.

Pokemon Go is designed for users to interact with analogical reality ($R^3$) through a digital reality layer ($DR^2$) by displaying a 2D experience on the screen of a mobile device that is a two-dimensional representation of a physical three-dimensional space enriched by a layer of digital information in two dimensions (f: $R^3 \rightarrow DR^2 \cap AR^2$).

The game allows for a fictional digital reality consisting of avatars and Pokemon (up to 802 fictional species of monsters) to take part of the two-dimensional visual representation of physical analogical reality and for that combination to be the "field of play" where the user interacts with $DR^2 \cap AR^2$ through his/her mobile device. As it currently proposes Facebook, the Pokemon Go AR is built on the screen of the mobile device through the action of the camera. The digital layer of AR is injected onto the digital representation of physical reality on a cartographic plane based on the geospatial location of the mobile device obtained through AGPS (assisted GPS, a combination of GPS, cell phone towers and WiFi networks signals).

The combination of Pokemon Go's interactive AR with an additional step in the currently limited AR proposed by Facebook could lead to the fourth generation social media, based on user social interaction with a digitally enhanced analogical reality.

Facebook conceives [7] three types of qualitative increases in the user experience based on AR: augmenting objects in analogical physical reality with additional digital information like putting a restaurant's reviews on their storefront to be displayed in the screen of the user's smartphone, tablet or computer when targeting the object with its camera; layering $VR^2$ objects onto analogical physical reality; and enhancing real objects (the screen digital display of real objects) with extra camera effects, like the currently available for static photos or recorded videos in Instragram, Snapchat or Whatsapp.

However, both Facebook and other social networks will provide the true advance towards fourth generation social media when the enhancement inherent to AR is injected onto the people's digital representation ($DR^2$) as well as onto the objects captured in $R^3$ by the cameras of mobile devices. In this way, social networks will implement a layer of People Augmented Reality (PAR) that will join what could be called the current Objects Augmented Reality (OAR).

Every person with a profile on social networks and the AGPS geolocation enabled on his/her mobile device can be an "object" with AR information popped up when that person is targeted by the camera of another user's mobile device working in AR enhanced camera effects mode.

The AR layer representing people with profiles in social networks can display other users basic or extended information of their profiles in social media, shared interests, images and, actually, everything that the user has defined by privacy criteria that can be shared through a PAR enriched layer. Likewise the person can share OAR information regarding his personal objects, such as if his/her car or his house connected to the Internet are for sale. Similarly, any company or business with profiles in social media might use OAR and PAR to, as their buildings or facilities are targeted with a camera of a mobile device, display on the screen of that device AR information about the company, its employees or any other data that may be shared in one or more layers of AR. Even public organizations can deploy AR information addressed to the people through a visible OAR layer on the screen of the users' mobile device through the camera effects.

Therefore, a fourth generation social network will be one that represents digital and two-dimensional ($DR^2$) physical reality ($R^3$) by enhancing it with a layer of OAR and another layer of PAR, resulting in f: $R^3 \rightarrow DR^2 \cap OAR^2 \cap PAR^2$.

## V. REFERENCES

[1] C. Thurlow, L. Lengen and A. Tomic, "Computer Mediated Communication: Social Interaction and Internet. Sage, London (2004).

[2] F.D. Kramer, S. Starr, L.K. Wentz (eds), "Cyberpower and National Security", National Defense University Press (2009).

[3] M. Graham, "Geography/Internet: Ethereal, Alternate Dimensiones of Cyberspace or Grounded Augmented Realities?, The Geographical Journal vol. 179, no. 2 (2012), pp. 177-188.

[4] https://en.wikipedia.org/wiki/Google_Glass, visited on 5/1/2017

[5] https://www.facebook.com/spaces, visited on 5/5/2017

[6] https://developers.facebook.com/blog/post/2017/04/18/Introducing-Camera-Effects-Platform/, visited on 5/6/2017

[7] https://techcrunch.com/2017/04/18/facebook-camera-effects-platform/, visited on 5/6/2017.